\documentclass[twocolumn,eqsecnum,preprintnumbers,nofootinbib]{revtex4}
\usepackage{amsmath,amssymb,amsfonts,times,graphicx}
\usepackage[ps2pdf,bookmarks=true,colorlinks,linkcolor=red,urlcolor=blue,citecolor=blue]{hyperref}
\usepackage[usenames]{color}
\usepackage{dcolumn}
\usepackage[normalem]{ulem}
\usepackage{enumerate}
\usepackage{epsfig}
\usepackage{yfonts}
\usepackage{bm}

\newcommand{\beq}{\begin{equation}}
\newcommand{\eeq}{\end{equation}}
\newcommand{\ba}{\begin{array}{ccc}}
\newcommand{\ea}{\end{array}}

\def\bea{\begin{eqnarray}}
\def\eea{\end{eqnarray}}

\begin{document}
\title{Interplay between short and long-range entanglement in symmetry protected phases}
\author{Brian Swingle}
\affiliation{Department of Physics, Harvard University, Cambridge MA 02138}

\date{\today}
\begin{abstract}
We study a variety of questions related to entanglement in symmetry protected phases, especially those introduced in arXiv:1106.4772 \cite{spt2}.  These phases are analogous to topological insulators in that they are short range entangled states with symmetry protected edge or surface states.  We show that the now standard bulk-edge correspondence relating the entanglement spectrum to the gapless edge spectrum holds for these phases as well.  We also consider the question of coupling these models to gauge fields or equivalently of introducing long range entanglement. We argue that this procedure yields models with perturbatively stable edge or surface states at a variety of interface types.  The non-onsite nature of the edge symmetry plays an important role in our considerations.
\end{abstract}

\maketitle

\section{Introduction}

The main theme of this investigation is the nature of boundary states in various kinds of quantum matter.  We now know of many examples of edge and surface states that are either completely robust (fractional quantum Hall states) \cite{PhysRevLett.48.1559,laughlin_wf,PhysRevB.25.2185,PhysRevB.43.11025} or are robust in the presence of certain symmetries (topological insulators) \cite{qsh_hgte,qsh_graph,3dti,ti_10fold}.  However, we still do not understand the general principles underlying the structure of edge states in 2d and surface states in 3d, not to mention to complexity of purely boundary phenomena.

The phases we study host a complex interplay of long and short range entanglement and promise to advance our understanding of the physics of surface states arising in topological quantum matter.  Such surface states are of experimental interest given the edge modes in quantum Hall bars \cite{PhysRevB.25.2185,PhysRevB.43.11025} and the surface states of three dimensional topological insulators \cite{3dti}.  Many of the examples of fractionalized phases for which we have a current experimental candidate host gapless states, either in the bulk or at the boundary (see for example \cite{PhysRevB.25.2185,PhysRevB.43.11025,3dti,qsh_hgte,qsh_graph,Yamashita04062010,PhysRevB.77.104413}).  This is not surprising given the general visibility of gapless states in experiment probes e.g. thermodynamics and transport.  Of course, we are not at this time proposing that the phases we consider here are directly relevant for current experiments, indeed, our exactly solvable models are quite artificial.  However, any advance in our understanding of gapless boundary states may be useful in future experiments given the prominence of these states in experimental examples of topological quantum matter.

We have a variety of interests in this work.  First, we wish to further understand the nature of symmetry protected edge states in systems with a unique ground state.  Key questions here include the conditions for symmetry protection, classification of edge states \cite{ti_10fold}, and the role of perturbations e.g. disorder. Second, we wish to understand further the relationship between the spectra of physical edges and the entanglement spectrum \cite{entspec}.  Recall that the entanglement spectrum is essentially the spectrum of the reduced density matrix of a subregion in the bulk of the material.  This connection is important not only because it provides a bulk method to identify symmetry protected phases, but also because it suggests how the edge classification feeds back into the bulk classification.  Third, we are interested in the extent to which short ranged symmetry protected phases can be strengthened by the addition of long range entanglement e.g. can we make these phases more robust in the presence of perturbations?  We now elaborate briefly on this last point.

In Refs. \cite{spt1,spt2} an exactly solvable model on a square lattice was given that has a unique ground state on any closed manifold and symmetry protected edge states on any open manifold.  The gaplessness of the non-chiral edge states is protected by an onsite $Z_2$ symmetry.  Either the edge breaks the symmetry or the edge is gapless.  While the symmetry is defined onsite i.e. the unitary generating the symmetry has the form $U = \prod_r U_r$, the natural ground state is a multipartite entangled state associated with plaquettes of the square lattice.  This tension leads to symmetry protection as codified by group cohomology and prevents the bulk $Z_2$ symmetry from being an onsite symmetry within the low energy edge degrees of freedom.  Indeed, the global symmetry operation requires operators that act at each effective edge site as well as between each neighboring pair of edge sites.  Importantly, this spatially extended symmetry structure is preserved under coarse graining.  However, this protection is only present as long as the onsite $Z_2$ symmetry is preserved.  Enter long range topological order.

Consider a long range entangled phase where the low energy excitations are described by an emergent $Z_2$ gauge theory as in Refs. \cite{kitaevexact,stringnet}.  Such a phase of matter need not possess any global symmetry, and we will assume no such symmetry.  However, the gauge theory description implies an emergent $Z_2$ gauge symmetry in the deconfined phase.  This is a somewhat strange symmetry that cannot be broken by any local physical perturbation, hence it is a natural candidate to protect the short ranged state described above.  To clarify what is meant, consider a gaped system with global symmetry $G$. An isolated subregion $R$ of that system can sometimes be labeled by a definite representation $\mathcal{R}$ of the group $G$ which is the total charge enclosed in $R$.  This label is robust to $G$ symmetric operations within $R$, but by adding $G$ non-invariant perturbations and by passing through a symmetry breaking state we can return to a symmetric state with any other charge $\mathcal{R}'$.  Now consider an emergent gauge theory with gauge group $G$.  We are sometimes still able to label isolated regions by a total charge (or more generally by anyon quantum numbers), but now no local perturbation can change this assignment.  Physically this is because there must exist an electric field labeled by $\mathcal{R}$ that connects $R$ with its environment.

We will argue that by coupling gauge fields to symmetry protected phases of matter we can produce systems with more robust edge and surface states.  As we discuss in more detail below, such states may not be protected, that is guaranteed to be gapless by some symmetry (since there is no physical symmetry), but they can be robust in the sense that no small perturbation can destroy them.  In this way we can trade symmetry based protection for increased perturbative robustness.

Independent of this work, gauging the symmetry of these symmetry protected phases has recently been considered in Ref. \cite{sptbraid} as a device to prove results about edge states in the ungauged model.  That work also considers interesting duality transformations between symmetry protected phases and long range entangled phases, and we comment briefly on the additional insights into our problem provided by this duality technology.

We will begin with some general motivational considerations before turning to a discussion of gapless boundary states.  Then we will elaborate on the bulk-edge correspondence in various symmetry protected phases before turning to the question of gauging these phases.  Our results include establishing a bulk-edge correspondence for the symmetry protected phases considered here and an analysis of the perturbative stability of edge states in gauged models.  We begin with some general motivating considerations in an effort to isolate the essence of symmetry protected edge states.

\section{Motivating considerations}
In many integer and fractional quantum Hall states, chirality and the bulk gap lead to an absolutely protected edge state, and in topological insulators we may have robust edge states so long as a certain symmetry is preserved.  Examples of states protected by time reversal include the quantum spin Hall state in two dimensions \cite{qsh_hgte,qsh_graph} and the $Z_2$ topological insulator in three dimensions \cite{3dti}.  These states have likely been realized in HgTe quantum wells as reported in Ref. \cite{hgte} and Bi2Te3 crystals as reported in Ref. \cite{arpesbi2te3} among others.  We can also consider classes of states called fractional topological insulators as in Ref. \cite{fti2d,fti3d1,fti3d2}.  Generally speaking, these states stand in relation to topological insulators in the same way that the integer and fractional quantum Hall effects are related.  One feature of these states is that the gapless boundary modes may be stable in an RG sense to perturbations that break the symmetry as in Refs. \cite{fti2dstable,exp_fti}.  This added robustness is in keeping with the general intuition that fractionalization enhances the perturbative stability of the state.  Such intuition arises from the understanding that fractionalization goes hand in hand with long range entanglement, a property that is often robust to local perturbations.  A general theme in the investigation of boundary states has been the topology of free fermion bandstructures, but this approach is not directly applicable to the symmetry protected states we consider.  Instead, as argued in Ref. \cite{spt2}, the low energy theory of the states in Ref. \cite{spt1,spt2} can be understood in terms of a $\theta$ term associated to a discrete symmetry.

$\theta$ terms and associated Wess-Zumino-Witten terms for continuous non-linear sigma models (NLSM) are well known in the theory of quantum magnetism (see Ref. \cite{topo_qmag_review} for a nice review).  For example, consider a $2+1$ dimensional bulk phase described by a gapped $SU(2)$ non-linear sigma model with a $\theta$ term.  If the space on which the system is defined has a boundary, then the bulk $\theta$ term, which is a total derivative when considering smooth field configurations, descends to a WZW term on the boundary of the manifold.  The action of this system is
\beq
S \sim \int dt\, dx\, \text{tr}\left[\partial^\mu U \partial_\mu U^\dagger \right] + S_{WZW}
\eeq
with
\bea
&& S_{WZW} \propto \int dt\, dx\, du\, \epsilon^{\mu \nu \lambda} \cr
&& \times \text{tr}\left[(U^\dagger \partial_\mu U)( U^\dagger \partial_\nu U)(U^\dagger \partial_\lambda U)\right].
\eea
The auxiliary space required to define the WZW term is here interpreted as the bulk $2+1$ dimensional gapped bulk.  Indeed, the integrand of the WZW term is nothing but the $2+1$ dimensional action for the $\theta$ term action.  The $1+1$ dimensional $SU(2)$ NLSM in the presence of a WZW term is an interesting system because, as was shown in Ref. \cite{nonabelbos}, there exists a stable RG fixed point.  Thus this system can enter a stable gapless phase at the boundary provided the $SU(2)$ symmetry is not broken explicitly.

For the purposes of our discussion here, there are two important ingredients to this story.  These ingredients arise from considering the two paradigmatic phases of a magnet, symmetry broken and paramagnetic.  The first ingredient is the fact that a continuous symmetry cannot break spontaneously in $1+1$ dimensions.  The second ingredient is the tendency of the WZW term to disfavor localized symmetric states e.g. a paramagnetic state.  Since this second piece of intuition is rather more vague, we briefly elaborate on it.

We can only write the WZW term as a local integral over spacetime provided we sacrifice the manifest symmetry of the Lagrangian.  When we make a global symmetry transformation on the Lagrangian, the manifestly symmetric terms are invariant without further thought, however, the WZW term changes by a total derivative.  The action is still invariant provided the space is closed, and this follows because we realize our WZW terms on the boundary of another bulk system (the boundary of a boundary is empty).  The fact that the Lagrangian is only invariant up to a total derivative is equivalent to the statement that the operator generating the symmetry transformation is not a product of terms acting on disjoint degrees of freedom.  In other words, we cannot simply remove degrees of freedom from the boundary while keeping the symmetry intact.  In this sense, a localized symmetric phase of the NLSM is in tension with the WZW term, for example, the quantum state must not be smoothly connected to a product state if the symmetry is preserved.

A simple example of this is phenomenon is provided by the spin-1/2 Heisenberg chain as shown in Ref. \cite{wzwspinchain}.  Of course, the spin rotation symmetry is defined onsite, but the gaplessness of this system may be understood in terms of an approximate ``superspin" symmetry that combines the Neel $\vec{N}$ and valence bond solid $D$ order parameters into a single $SU(2)$ matrix $U = D + i \vec{N}\cdot \sigma$ (we impose the constraint $D^2+ |\vec{N}|^2=1$).  The low energy theory is then a NLSM for the field $U$ with a WZW term at level one.  This theory possesses a stable RG fixed point and hence a gapless phase.  Now in accord with our general comments above, it should not possible to realize the larger superspin symmetry in an onsite fashion.  Define $M_i = N_i$ and $M_4 = D$ and write the $SU(2)$ WZW term as
\bea
&& S_{WZW} \propto \int dt\, dx\, du\, \epsilon^{IJKL}\epsilon^{\mu \nu \lambda} \cr
&& \times M_I \partial_\mu M_J \partial_\nu M_K \partial_\lambda M_L .
\eea
This term may be cast as a two dimensional action at the cost of losing manifest symmetry.  Set $M_i = m_i$ and $M_4 = \sqrt{1-m^2} \approx 1$ where the $m_i$ should be regarded as infinitesimal and the symmetry acts by acts $m_i =\rightarrow m_i + \delta_i$.  Expanding the WZW term about this classical configuration gives
\beq
S_{WZW} \propto \int dt \,dx \,\epsilon^{\mu \nu} \epsilon^{i jk} m_i \partial_\mu m_j \partial_\nu m_k
\eeq
and so we see that $\delta S_{WZW}$ is
\beq
\delta S_{WZW} = \delta_i \int dt \,dx \,\epsilon^{\mu \nu} \epsilon^{i jk} \partial_\mu m_j \partial_\nu m_k
\eeq
which is only zero up to a total derivative.  Microscopically the non-onsite nature of the symmetry follows because the VBS state consists of entangled neighboring spins while the Neel state is a product and no unitary formed from a product of onsite unitaries can connect these two states.  We note that the appearance of an enlarged symmetry at the low energy fixed point, here an enhanced $SU(2)_L \times SU(2)_R$ symmetry, also appears to be a common feature of symmetry protected edge states.

Now as we said, a discrete generalization of $\theta$ and WZW terms were considered in Ref. \cite{spt1} and argued to lead to a gapless edge.  Unfortunately, we do not yet have much understanding of this phenomenon from a field theory point of view.  Nevertheless, early results in Refs. \cite{spt2,boson_iqh1,boson_iqh2} suggest the following picture at least in $d=2$: the edge theory is non-chiral but the symmetry is represented chirally on the edge modes.  In other words, the left movers, say, carry the conserved charge while the right movers do not.  Hence any backscattering terms that would gap the edge are forbidden by symmetry.  Furthermore, the basic intuition described above about non-onsite boundary symmetry also carries over, and we will see below the non-onsite character of the boundary symmetry explicitly below.  A possibility that must be considered in the discrete case is spontaneous symmetry breaking at the edge (this is also an issue in continuous models in higher dimension).  This still leads to ``gapless" states at the edge but now associated with the order parameter direction in the symmetry broken state.

\section{Boundary states and non-onsite symmetries}

Let us now understand the role of the non-onsite action of the symmetry $Q$ in protecting the low energy edge degrees of freedom.  Onsite means that $Q$ may be written as a product $Q = \bigotimes_r Q^{(r)}$ with $(Q^{(r)})^\dagger = (Q^{(r)})^{-1}$ for all $r$.  Examples include rotation about the $z$-axis in spin systems, $Q(\theta) = \bigotimes_r \exp{(- i \theta S^z)}$, and phase rotations in boson systems, $Q(\theta) = \bigotimes_r \exp{(-i\theta n_r)}$ ($n_r$ is the boson number operator).  In general, we would require that $Q$ be defined as a sitewise product of unitary representations of some group.  We will be interested in precisely those cases where the edge symmetry is not onsite in the above sense (nor should it be short range equivalent to an onsite symmetry).

Consider a bulk gapped symmetry protected phase with a physical boundary.  Suppose that the ground state is invariant under the symmetry and fully gapped in the gap.  Let us further imagine that at the edge there are a set of low lying states, not necessarily gapless, but lying much below the bulk gap.  For the phases considered here, we can construct exactly solvable models where the edge is associated with a degenerate manifold of states.  Adding perturbations yields the low energy manifold of interest.  We wish to understand under what conditions this low energy manifold is guaranteed to form a gapless state.  Let $Q$ represent the low energy part of the action of the bulk symmetry on the edge states.  We will be much more explicit about this construction when we give some solvable models.  The key issue will turn out to be whether $Q$ has a low energy part that is short range equivalent, meaning equivalent under a short range unitary, to an onsite symmetry.

To get some intuition, suppose $Q$ is related via a finite depth circuit $V$ to an onsite symmetry $Q = V(\bigotimes_{r \in E} Q^r)V^{-1}$.  Then clearly the edge can be in a gapped and short ranged state.  Indeed, take any gapped onsite Hamiltonian $H_E = \sum_{r\in E} h_r$ with $[h_r,Q^r] =0$ and consider its ground state $|\psi \rangle$.  We can always find such a Hamiltonian by slightly coarse graining the system (although translation invariance may need to be broken).  The state $V |\psi \rangle$ is clearly invariant $Q$, it is short range entangled, and is the ground state of a local Hamiltonian with a finite gap.  What happens if we cannot relate $Q$ to any onsite symmetry generator via a finite depth quantum circuit?

Let us consider a simple $Z_2$ example.  We use the notation $Z$ and $X$ for the Pauli operators.  Let $Q$ be defined on N spin-1/2 degrees of freedom labeled by $Z_r=\pm 1$ with action
\beq
Q = \tilde{Q} \prod_r X_r
\eeq
where
\beq
\tilde{Q} = \exp{\left[i \pi \sum_r \frac{(1-Z_r)(1-Z_{r+1})}{4}\right]}.
\eeq
This symmetry as defined is clearly not onsite, but it is also the case that it is not short range equivalent to an onsite symmetry (this will be clear later).  As an additional comment, this symmetry may at first seem quite special, but from a renormalization group perspective it is fairly general.  In the extreme infrared limit, we should be able to express any symmetry operation, even non-onsite symmetries, in terms of actions on neighboring spins.

Let us now introduce a family of Hamiltonians $H(g)$ that all commute with $Q$.  To prove that no ground state of $H(g)$ can be symmetric and short range entangled, we make two general assumptions:
\begin{itemize}
\item{[Invariance]} The ground state $|\psi(g)\rangle$ is unique, gapped, and invariant under $Q$.
\item{[Local operators]} There are local operators $Z_r$ satisfying $QZ_r = - Z_r Q$ (here this follows by definition of $Q$, but in general we will want to restrict attention to $Q$ for which this is true).
\end{itemize}
Using these assumptions we can prove the absence of symmetric short ranged states.

Let $g^\star$ correspond to a state deep within a gapped short ranged entangled symmetric phase and write $|\psi(g^\star)\rangle = \sum_{z=\{z_r\}} c(z) |z_1 ... z_N\rangle$.  The invariance of $|\psi(g^\star)\rangle $ implies the equation
\beq\label{tildeQ}
\tilde{Q}(z) c(-z) = c(z)
\eeq
for all configurations $z$.  An immediate corollary is that $|c(z)| = |c(-z)|$.  Since $g^\star$ represents a point deep in a short ranged entangled state, we may write the wavefunction as
\beq
c(z) = \prod_r f(z_r,z_{r+1}),
\eeq
and furthermore, given the extreme short ranged nature of the states, we need only consider a short chain of four sites (the minimal number).  The extension to larger chains will be obvious.  We also assume that $f(z,z') = f(z',z)$.  Now consider the configuration $1_1 1_2 -1_3 -1_4$ where the subscripts indicate the site and we have periodic boundary conditions.  Eq. \ref{tildeQ} implies that
\bea
&& f(1,1)f(1,-1)f(-1,-1)f(1,-1) \cr
&& = - f(1,1)f(1,-1)f(-1,-1)f(1,-1).
\eea
Clearly this can only be true if some of the $f$s are zero.

Suppose $f(1,-1) =0$, then it follows that only configurations of the form $1...1$ and $0...0$ have non-zero amplitude i.e. the state is a ferromagnetic cat state (in order to have invariance) or otherwise breaks the symmetry.  Suppose $f(1,1) = 0$.  The condition that $|c(z)| = |c(-z)|$ immediately implies that $f(-1,-1)=0$ and hence that the state is anti-ferromagnetic.  The same result obtains if we suppose initially that $f(-1,-1) =0$.  Thus we have shown that a symmetric short ranged state is incompatible with the non-onsite symmetry $Q$.  In fact, we know slightly more, since there are long range correlated states with $c(z)$ given as above e.g. a state where $c(z) = e^{- \beta_c h_{I}(z)/2} $ with $h_{I}$ the classical Ising Hamiltonian and $\beta_c$ the critical temperature.  These kinds of states are also ruled out despite their power law correlations for some observables.

Our proof is certainly more elementary than the sophisticated proof using matrix product states in Ref. \cite{spt1}, but we can also check that in other simple cases the same logic applies.  We are currently developing this more elementary approach, which focuses directly on the non-onsite nature of edge symmetry, as an alternative to the group cohomology logic \cite{inprep}.  Note also that our wavefunction is equivalent to a matrix product state with matrices given by $A^m_{\alpha \beta} = \sqrt{g(\alpha, m) g(m, \beta)} $ ($m$ the physical index and all indices running over $\{1,-1\}$), and hence we make the same assumption for the form of the edge wavefunction as in Ref. \cite{spt1}.  One of the main virtues of our approach, which focuses directly on the non-onsite symmetry, is that it is generalizable to higher dimensions \cite{inprep}, a situation for which no formal proof of gaplessness for non-onsite symmetries yet exists.

Once we consider the possibility of spontaneous symmetry breaking, the generic statement is that the edge spectrum must always have low lying states.  These low lying states may either be due to the presence of gapless edge modes or to the presence of degenerate ground states coming from spontaneous symmetry breaking.  Having recapitulated our understanding of the protected low lying nature of the physical edge spectrum, we turn now to our second primary interest, the bulk entanglement spectrum.

\section{Entanglement spectrum in symmetry protected phases}

Recall that the entanglement spectrum of a region $R$ is defined using the reduced density $\rho_R = \mbox{tr}_{\bar{R}} (\rho)$ via the ansatz $\rho_R = e^{-K_R}$.  The spectrum of $K_R$ is called the entanglement spectrum of $R$ when the state $\rho$ is the ground state.  Note that the word ``entanglement" is only appropriate when the state $\rho$ is pure, otherwise not all the information encoded in the spectrum of $K_R$ describes entanglement.  The entanglement entropy of $R$ is simply the thermal entropy of the Hamiltonian $K_R$ at temperature one.  More generally, an unentangled state $\rho$ is described by an entanglement Hamiltonian with a single zero eigenvalue and an infinite gap to all other states i.e. $e^{-K_R}$ is a projector of rank one.  On the other hand, if $H_R$ is gapless then the state $\rho$ is highly entangled between $R$ and $\bar{R}$ ($R$s complement).  Thus the entanglement spectrum provides an interesting visceral measure of entanglement as well as permitting the computation of more precise measures such as the Renyi entropies $S_{\alpha}(R) = (1-\alpha)^{-1} \log{(\mbox{tr}(\rho_R^\alpha))}$.  Now we ask how the entanglement spectrum looks in a symmetry protected phase of the type we have been considering.

Consider a bulk Hamiltonian $H_B$ with gapped ground state $|\psi_B \rangle$.  Suppose the Hamiltonian commutes with an onsite symmetry $Q_B$.  Again, onsite means that $Q_B$ may be written as a product $Q_B = \bigotimes_r Q^{(r)}$ with $(Q^{(r)})^\dagger = (Q^{(r)})^{-1}$ for all $r$.  In general, we simply require that $Q_B$ be defined as a sitewise product of unitary representations of some group.  Suppose also that the ground state is also invariant under the symmetry $Q_B$: $Q_B |\psi_B \rangle = | \psi_B \rangle$.

If we now consider a subregion $R$, we find that the state of $R$ is still invariant under the symmetry in the sense that
\beq
\rho_R = \mbox{tr}_{\bar{R}}(\rho) = \mbox{tr}_{\bar{R}}(Q_B \rho Q_B^{-1}) = Q_R \rho_R Q_R^{-1}
\eeq
where $Q_R = \bigotimes_{r\in R} Q^{(r)}$.  In other words, we have $[K_R,Q_R]=0$.  Note that it is crucial for this result that we trace out degrees of freedom on which the symmetry acts in the standard way i.e. we don't split the symmetry action across the boundary.  The boundary law for entanglement entropy plus the bulk gap suggests that $K_R$ should be interpretable as a local Hamiltonian for a physical edge.  Let $|\{m_r\}\rangle $ be a basis for the states in $R$ with finite entanglement energy (in practice we can add a high entanglement energy cutoff so that this basis is much smaller than the full basis for region $R$).  Suppose that the labels $\{m_r\}$ are distinguished by local changes of configuration confined to the boundary of $R$ as suggested by the finite gap and correlation length. We may thus ask how $Q_R$ acts on the states $|\{m_r\}\rangle $.  This action is given by a new matrix $Q$ defined by
\beq
Q_R |m\rangle = \sum_{m'} Q_{m m'} |m'\rangle.
\eeq
Our key result is that $Q_{m m'}$ is non-onsite with respect to the local $m$ degrees of freedom and is short range equivalent to the symmetry action on physical edge states.  In other words, the entanglement eigenstates support the same non-onsite symmetry action as the physical edge states.  Since the entanglement Hamiltonian $K_R$ commutes with a non-onsite symmetry and is local at the edge, our previous arguments imply that its spectrum is necessarily gapless e.g. the entanglement Hamiltonian is truly gapless or spontaneously breaks the symmetry.

We will show this non-onsite structure explicitly below for the $Z_2$ model, however, the result is quite general and follows directly from the form of the fixed point wavefunctions in Ref. \cite{spt1,spt2}.  Because those wavefunctions are direct products we can explicitly construct the entanglement boundary degrees of freedom $m$ and the entanglement Hamiltonian is trivially local.  The cohomology classification then implies that such a local Hamiltonian symmetric under $Q_R$ must have edge states.  A similar result is immediately visible in the field theory for the non-linear sigma models we were considering above.  The results of Ref. \cite{entspec6} imply that the entanglement spectrum of a relativistic NLSM, such as a NLSM with bulk $\theta$ term, is equivalent in its universal properties to the physical edge spectrum.  But as we already described above, the physical edge of such a gapped bulk phase with $\theta$ term is a NLSM with a WZW term and hence (at least in $1+1$ dimensions) is gapless.  Thus the entanglement spectrum is also gapless like the physical edge spectrum and the action of the symmetry generator is again non-onsite.

We can also use the recent results of Ref. \cite{cs_spt_classify} to give an alternate proof of the bulk-edge correspondence for some phases in the cohomology classification considered in Refs. \cite{spt1,spt2}.  The proof proceeds exactly as in the previous paragraph.  Ref. \cite{cs_spt_classify} provides relativistic field theories taking the form of Chern-Simons theories for several classes of symmetry protected phases.  The results of Ref. \cite{entspec6} thus immediately imply that the bulk-edge correspondence holds for these systems.  We must only choose symmetric boundary conditions for the Rindler space cutoff in \cite{entspec6} to demonstrate the gapless edge modes.  Symmetry breaking is also a possibility, but as we described above the entanglement spectrum will still possess low lying states. Note also that this proof is quite different from the arguments above based on a lattice model formulation.  Indeed, our arguments above apply to all the cohomology phases in Ref. \cite{spt2}, but the field theory approach still only works for a limited number of these models with known low energy descriptions.  This is clearly a promising direction for further progress.

We have seen that the entanglement Hamiltonian can naturally function as a local Hamiltonian for the system with a physical edge.  Furthermore, the bulk gap in such an entanglement Hamiltonian goes to infinity away from the edge.  This last requirement is crucial to obtain a boundary law for the entanglement entropy since any finite gap bulk system at finite temperature will give a volume contribution to the entropy.  As long as we make this stipulation that the gap go to infinity in the bulk, it seems consistent to take the entanglement Hamiltonian to basically be any local Hamiltonian for the topological state with edge.  However, we must be quite careful here.

If the actual Hamiltonian of the bulk material is translation invariant with a translation invariant ground state, then despite the appearance of a non-translation invariant operator $K_R$ in the exponential, the state $e^{- K_R}$ is actually translation provided the temperature is kept at one.  On the other hand, the generalized operator $e^{- \beta K_R}$ is not generically translation invariant for $\beta \neq 1$.  Nevertheless, these operators are interesting since they control the Renyi entropies $S_\alpha(R) = (1-\alpha)^{-1} \log{(\text{tr}(\rho_R^\alpha))}$.  By sending $\beta \rightarrow \infty$ we can project onto the entanglement Hamiltonian's ground state which we have argued should be the ground state of the physical system with edge.  In particular, if the edge is gapless, like the edge of a quantum Hall state, then this entanglement ground state will possess long range correlations at the edge.  This is another quite intriguing feature of the reduced density matrix in topological phases: where the original density matrix $e^{- K_R}$ encodes a mixed state with decaying correlations, hidden within this density matrix is a pure state with long range correlations.

What happens when the entanglement Hamiltonian has a symmetry breaking ground state?  Recall that even in a symmetry broken phase, the spectrum is symmetric, but there is usually a breakdown of ergodicity that prevents the full manifold of symmetric states from being explored.   However, the true ground state e.g. a cat state will also possess long range correlations.  This in this case we see that the $\beta \rightarrow \infty $ limit still produces a long range correlated state.  We can use this observation to develop an interesting argument for a gapless surface in three dimensional systems.  Recall that in two spatial dimensions a discrete symmetry breaking phase can survive to finite temperature.  If the surface of a three dimensional symmetry protected phase generically breaks the symmetry, then the entanglement Hamiltonian must similarly have a symmetry broken ground state.  Normally we think of a breakdown of ergodicity in such a state, but if we use the formal Gibbs state, then cluster decomposition is violated i.e. correlators do not factor at large distances.  In particular, there are long ranged correlations in such a state, but long ranged correlations are incompatible with a gapped bulk state.  Thus it must be that the thermal system described by $\exp{(-K_R)}$ is always in a paramagnetic phase.  Either $K_R$ has a symmetric low temperature phase (which is necessarily gapless) or the ground state of $K_R$ breaks symmetry and the ``entanglement temperature" must always be greater than the ordering temperature.  If the entanglement Hamiltonian does have a symmetry breaking ground state, then we can detect this by taking the limit $\beta \rightarrow \infty $ for the operator $e^{-\beta K_R}$ as above.  Thus we see again that one simple characterization of the bulk of a symmetry protected state is that the operator $\lim_{\beta\rightarrow \infty} e^{-\beta K_R}$ always contains long range correlations due either to symmetry breaking or gapless states in the entanglement spectrum.

\section{A 2d example: the gauged CZX model}
Having given a general picture of the protected edge modes at a physical boundary as well as the bulk-edge correspondence for the entanglement spectrum, we would now like to turn to the question of edge stability in the presence of long range entanglement.  The simplest setting where we can carry out the process of coupling to gauge fields is in the non-trivial $Z_2$ phase described in Ref. \cite{spt1}.  This phase is paramagnetic just like the large transverse field limit of the Ising model, but the phases are nevertheless distinct as demonstrated by the group cohomology arguments of Ref. \cite{spt1,spt2}.  We will explain how to gauge this model, which is known as the CZX model (for reasons that will become clear), and then later discuss more general constructions.  We will also give a preliminary discussion of the edge structure at the end of this section.

\begin{figure}
  \centering
  \includegraphics[width=.3\textwidth]{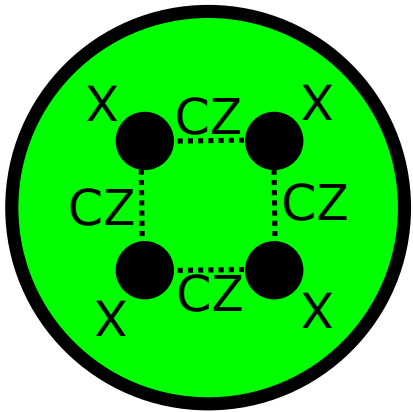}
  \caption{The green dot represents a single site on the square lattice.  The onsite $Z_2$ symmetry acts by spin flip on each qubit and by a controlled Z operation on each neighboring pair of qubits.  These two steps commute with each other and square to one.
  }
  \label{fig:czxsite}
\end{figure}
Let us begin with a review the ungauged CZX model.  The CZX model is defined on a square lattice with four qubits per site (see Ref. \cite{spt1}).  These qubits carry an onsite $Z_2$ symmetry whose action is given
\beq
Q = (X \otimes X \otimes X \otimes X)(CZ\otimes CZ \otimes CZ \otimes CZ)
\eeq
where $X$ is the single qubit Pauli and $CZ$ is a controlled $Z$ gate acting on pairs of qubits.  The symmetry structure is shown in Fig. \ref{fig:czxsite}.  Note the peculiar ``extended" structure of the symmetry action.  In the computational basis the action of $Q$ is
\beq
Q | a b c d\rangle = (-1)^{ab+bc+cd+da} |a+1\,b+1\, c+1\, d+1 \rangle.
\eeq
A simple computation shows that $Q^2=1$.  It is convenient to diagonalize $Q$ in order to classify onsite states as $Z_2$ charged or uncharged.  The eigenstates of $Q$ are shown in Table 1.
\begin{table}\caption{Eigenstates of the $Z_2$ symmetry $Q$}
\begin{tabular}{|l|c|c|c|}
  \hline
  states & $Q$ & number & name \\
  \hline
  $(1/\sqrt{2})(|0000\rangle \pm |1111\rangle)$ & $\pm 1$ & $2$ & $|\pm\rangle $ \\
  $(1/\sqrt{2})(|1000\rangle \pm |0111\rangle) (+ \text{trans.})$ & $\pm 1$ & $8$ & $|1,\pm\rangle (+ \text{trans.}) $ \\
  $(1/\sqrt{2})(|1100\rangle \pm |0011\rangle)$ & $\mp 1$ & $2$ & $|12,\mp \rangle $ \\
  $(1/\sqrt{2})(|0110\rangle \pm |1001\rangle)$ & $\mp 1$ & $2$ & $|23, \mp \rangle $ \\
  $(1/\sqrt{2})(|1010\rangle \pm |0101\rangle)$ & $\pm 1$ & $2$ & $|13, \pm \rangle $ \\
  \hline
\end{tabular}
\end{table}

\begin{figure}
  \centering
  \includegraphics[width=.4\textwidth]{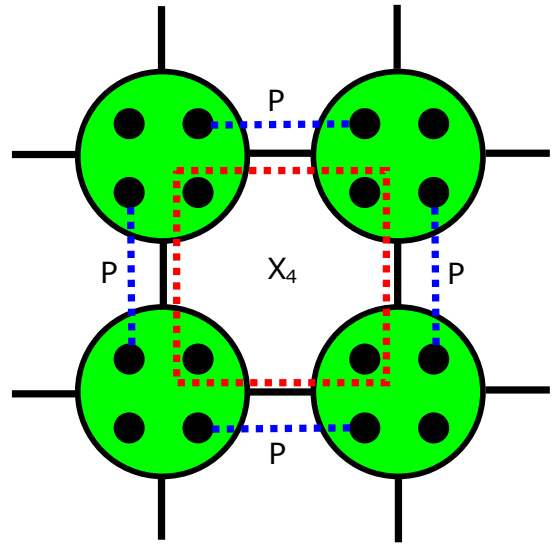}
  \caption{The plaquette term of the CZX Hamiltonian involves $12$ of the $16$ qubits in a plaquette.  The blue dashed lines indicates a a projector $P$ involving those qubits.  The red dashed box indicates the four qubits involved in the $X_4$ operator.  See the text for definitions.
  }
  \label{fig:czxham}
\end{figure}
The Hamiltonian of the CZX model is a sum of plaquette operators $H_p$ acting on twelve of the sixteen qubits involved in a plaquette:
\beq
H_{CZX} = \sum_p H_p.
\eeq

The structure of the interaction is shown in Fig. \ref{fig:czxham}.  The operator $X_4$ is
\beq
X_4 = |0000\rangle \langle 1111 | + |1111\rangle \langle 0000|
\eeq
while the projecters $P$ are
\beq
P = |00 \rangle \langle 00| + |11\rangle \langle 11|.
\eeq
Focusing on a single four qubit site, we must consider operators of the form
\beq
O_{abc,a'b'c'} = |abc\rangle \langle a'b'c'| \otimes I
\eeq
where the identity $I$ acts on the outer corner.  In fact, when dealing with the CZX Hamiltonian we need only consider operators with $a=a'$, $c=c'$, and $b=b'+1$ (all arithmetic is mod $2$).  We label these operators $O_{abc}$ which is short for $O_{abc,ab+1c}$.  The action of $Q$ on these operators is given by
\beq
Q O_{abc} Q = (-1)^{a+c} O_{a+1\,b+1\,c+1}.
\eeq
A single plaquette term in the CZX Hamiltonian may be reexpressed as
\beq
H_p = - \sum_{a_1 a_2 a_3 a_4 b} O_{a_1 b a_2}^{(1)} O_{a_2 b a_3}^{(2)} O_{a_3 b a_4}^{(3)} O_{a_4 b a_1}^{(4)}
\eeq
where the $(r)$ superscript labels the site.  When acting with $Q^{(1)}Q^{(2)}Q^{(3)}Q^{(4)} $ on this operator we find that all the phases cancel and the terms are simply permuted about.  This establishes the symmetry of the CZX Hamiltonian under global symmetry transformations.  We also note that the different plaquette terms multiply to zero $H_p H_{p'} = 0$

\begin{figure}
  \centering
  \includegraphics[width=.4\textwidth]{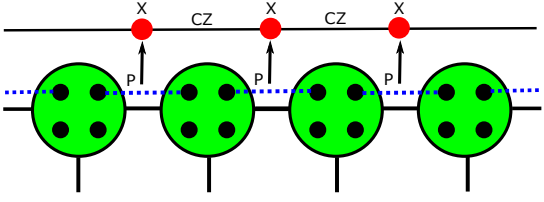}
  \caption{Here we show a physical edge in the CZX Hamiltonian together with the effective one dimensional description above it.  The blue dashed lines represent qubits constrained to the subspace $\{|00\rangle, |11\rangle\}$ due to a bulk projector $P$ (from the plaquette term immediately below the edge).  These two states are mapped to the effective site represented by the red dot with states $|\tilde{0}\rangle = |00\rangle$ and $|\tilde{1}\rangle = |11\rangle$.  As the figure indicates, in terms of these effective low energy edge states, the $Z_2$ symmetry is not onsite.  We must act with X on the tilde variables at each red site but also with CZ operators between pairs of sites.
  }
  \label{fig:czxedge}
\end{figure}

The ground state of the CZX Hamiltonian is beautifully simple.  We simply place the four inside corner qubits of each plaquette, those in the red dashed box of Fig. \ref{fig:czxham}, into the state
\beq
|0000\rangle + |1111 \rangle
\eeq
and tensor over all plaquettes.  Although the ground state is unique on a torus, the system has non-chiral edge states so long as the $Z_2$ symmetry is preserved. The extended nature of symmetry action on the edge, as mentioned in the introduction, is shown in Fig. \ref{fig:czxedge}.  The application of group cohomology ideas allows to prove the following result: the ground state of the edge cannot be written as a matrix product state with finite bond dimension in any phase where the symmetry action is preserved and described by a non-trivial element of the group cohomology.  This leaves two alternatives: either the bond dimension is infinite (gapless edge) or the symmetry is broken.

Unfortunately, it seems that in the $Z_2$ non-trivial phase the edge must be fine tuned to be gapless.  In other words, the edge generically breaks the symmetry.  We describe this in more detail below.  Note that this implies that two tunings are necessary to reach a gapless edge: we must impose the symmetry and then further tune one relevant operator to zero at the edge.  This defect is not present in the $Z_N$ models we briefly consider below or more generally.

We now wish to add strings to the model so that the resulting phase has long range topological order as in Refs. \cite{kitaevexact,stringnet}.  We can always rewrite the operators $O_{abc}$ in terms of the diagonal states in Table 1.  This is convenient if we want to gauge the symmetry.  To each link $rr'$ of the square lattice we associate a $Z_2$ variable $\tau_{rr'}^z$.  Consider a local $Z_2$ transformation
\beq
Q(r) = \prod_r (Q^{(r)})^{x_r}
\eeq
specified by $\{x_r\}$ ($x_r \in \{0,1\}$).  Under this transformation we take $\tau^z_{rr'} \rightarrow \tau^z_{rr'} (-1)^{x_r +x_{r'}}$.  This establishes the $\tau^z$ variables as gauge fields.  $\tau^x$ is analogous to an electric field operator, and we interpret $\tau^x=\pm 1$ as the absence or presence of a $Z_2$ field line or ``electric string".   Using the variable $\tau^z_{rr'}$ we may ``parallel transport" any quantity transforming under $Q^{(r')}$ to a quantity transforming under $Q^{(r)}$.  More generally, using strings of the $\tau$s we can parallel transport a $Z_2$ charged quantity from one site to any other site.  To construct the Hamiltonian of the gauged model, we first modify the plaquette operators of the CZX model.  We express the plaquette operators in terms of the states in Table 1 for the four sites surrounding a plaquette.  This is a tedious but straightforward exercise.  Then, for each term in the resulting expression, we append factors of $\tau^z_{rr'}$ so that the resulting operator is only charged under the local symmetry of a single site in the plaquette, say the upper left corner.  Because the original operator was invariant under the global transformation, the new operator will be invariant under all local transformations.

As an example of the first step in this procedure, consider the operator $O^{(r)}_{000}$ on site $r$.  This operator is
\beq
O_{000} = | 0 0 0\rangle \langle 0 1 0|(|0\rangle \langle 0| + |1\rangle \langle 1|).
\eeq
In terms of the states in Table 1 it may be written
\bea
&& \frac{1}{2}( |3,+\rangle + |3,-\rangle)(\langle 3,+| + \langle 3,-|) \cr
&& + \frac{1}{2} ( |13,+\rangle - |13,-\rangle)(\langle 13,+| - \langle 13,-|).
\eea
The $\pm$ signs dictate how the states transform under $Q^{(r)}$ so that
\bea
&& Q^{(r)} O_{000} Q^{(r)} = \frac{1}{2}( |3,+\rangle - |3,-\rangle)(\langle 3,+| - \langle 3,-|) \cr
&& + \frac{1}{2} ( |13,+\rangle + |13,-\rangle)(\langle 13,+| + \langle 13,-|).
\eea
Effectively, $+$ states are $Z_2$ neutral while $-$ states are $Z_2$ charged.  To parallel transport $O^{(r)}_{000}$ so that it transforms under $Q^{(r')}$ instead of $Q^{(r)}$ we modify it as follows
\bea
&& \frac{1}{2}( |3,+\rangle + \tau^z_{rr'} |3,-\rangle)(\langle 3,+| + \langle 3,-| \tau^z_{rr'}) \cr
&& + \frac{1}{2} ( |13,+\rangle - \tau^z_{rr'} |13,-\rangle)(\langle 13,+| - \langle 13,-| \tau^z_{rr'}).
\eea
This exercise must be carried out for the all $O_{abc}$ operators.

The full Hamiltonian is a sum of three terms.  The first term is the modified plaquette term of the CZX model.  The second term is a ``magnetic" term for the gauge fields given by
\beq
- \sum_{p} \prod_{rr' \in p} \tau^z_{rr'}
\eeq
where the sum is over all plaquettes $p$ and the product is over the four links in a given plaquette.  The third term is an ``electric" constraint term that takes the form
\beq
- \sum_r Q^{(r)} \prod_{r' \in nn(r)} \tau^x_{rr'}
\eeq
where the product is over all sites next to site $r$.  This term simply says that the number of electric field lines entering site $r$ must be equal to the $Z_2$ charge of that site (mod 2).  One may easily check that all three terms commute with each other.  The only non-trivial checks involve the electric term, but commutativity follows because the other two terms either move a charge and create a string segment or create a closed string.

The ground state of this model is simple enough.  We take the $Z_2$ plaquette state described above and tensor it with a state of the $\tau$ variables given by $\tau^z= 1$ on all links.  We then project onto the constraint given by the electric term.  This projection attaches $Z_2$ charges to the ends of open strings.  The resulting state is a sum over all string states with strings required to end at sites with non-trivial $Z_2$ charge as dictated by the bosonic wavefunction.

The model we have defined represents the extreme deconfined limit of gapped $Z_2$ charged bosons coupled to a $Z_2$ gauge field.  To obtain the ``true" gauge theory, we should impose the electric term as an actual constraint instead of energetically.  We take this route below for convenience, but the we should bear in mind that this is not necessary since the gauge structure is totally robust.  The low energy theory is then deconfined $Z_2$ gauge theory coupled to a $Z_2$ topological NLSM with $\theta$ term as discussed above i.e. a gauged topological NLSM.  At an interface between two deconfined phases, one with the $\theta$ term and one without, a lower dimensional WZW term will appear.  As an aside, in the high energy literature there are subtleties associated with gauging a WZW term, but we have not yet understand the role of these obstructions in our setting.  We already noted that the edge theory tends to spontaneously break the $Z_2$ symmetry.  However, as we will argue in more detail below, the emergent gauge structure prevents us from adding terms that explicitly break the $Z_2$ symmetry within the deconfined phase.  Thus it appears that while a gapless edge still requires fine tuning, that tuning is less than in the ungauged model since the symmetry is now effectively unbreakable.  Furthermore, the ungauged 2d models with $Z_N$ ($N>2$) symmetry (and potentially more general groups) have a stable gapless phase over a finite parameter regime.

At low energies this model is essentially equivalent to the original gapped boson model plus gapped $Z_2$ vortices.  But importantly, because the local $Z_2$ symmetry is an emergent gauge symmetry, no physical perturbation can break the $Z_2$ symmetry.  Alternatively, stability of the phase implies the existence of a renormalized gauge charge that always generates local symmetries.  So long as such a renormalization procedure cannot change the emergent cohomological structure, the special edge states should remain.  For example, the ``global" gauge transformation given by $x_r = x$ leaves all gauge fields invariant.  This transformation has the same spatially extended action at an edge as before, and since this structure is robust under renormalization, fluctuations of the gapped gauge fields should not be able to modify it.  We now discuss a few different types of edges in these theories.

Consider first a boundary between two deconfined phases, one with symmetry protected short range topological order and one without.  Since nothing is happening to the gauge field, and since the symmetry cannot be broken by terms in the physical Hamiltonian, we expect the system will have robust non-chiral edge states.  However, we emphasize that these states are not protected. It is also possible for these edge states to spontaneously break the gauge symmetry and thus enter a Higgs phase.  In this case, the gauge theory is locally screened, the emergent symmetry is lost, and the protected edge states may disappear.

Now consider a boundary between the exotic deconfined phase and vacuum.  Here the question of edge states is more delicate. We might expect them to become gapped due to proximity to a confined state since the emergent symmetry is effectively destroyed in such a phase.  For example, sometimes a confined state is smoothly connected to a Higgs phase of the gauge theory, and as we argued above, proximity to a Higgs condensate will destroy the edge states.  However, this argument is heuristic only since the vacuum, or really, open boundary conditions in a lattice model, is not obviously the same as a confined state.  Let us ask what happens in the lattice model, first in the ungauged model and then with strings.

If the CZX model is given a boundary by removing all the plaquette terms above some line, the model now has a large ground state degeneracy.  The projectors coming from the plaquette term just below the boundary effectively set neighboring boundary spins equal as shown in Fig. \ref{fig:czxedge}.  The resulting degenerate manifold is already an indication of gaplessness, but to see a dispersion we must add extra perturbations to the edge.  These perturbations must be consistent with the action of the edge symmetry shown in Fig. \ref{fig:czxedge}.  The simplest such perturbation is an Ising coupling of the for $\tilde{Z}_r \tilde{Z}_{r'}$ which commutes with the symmetry.  Another possibility is $\tilde{X}_r + \tilde{Z}_{r-1} \tilde{X}_r \tilde{Z}_{r+1}$.  The unusual structure here is a consequence of the extended nature of the symmetry.  For example, we already see that we cannot simply add a strong transverse field.  We must also couple in an additional three spin term.  This is precisely the unusual edge symmetry preventing us from reaching a short ranged symmetric phase.

On general grounds we know that the edge Hamiltonian cannot be tuned to point where the ground state is a gapped symmetric state.  This leaves two possibilities.  Possibility 1: the edge possesses a gapless phase analogous to the continuous group case.  Possibility 2: the edge always breaks the symmetry unless we fine tune.  Possibility 1 requires the absence of all relevant operators in the gapless theory.  In the conventional Ising model the transverse field can tune through a phase transition, however we have seen this term is not allowed on its own.  Suppose the theory does possess a relevant operator.  It must be the case that adding this relevant operator to the fixed point action leads to symmetry breaking regardless of the sign of the coupling.  For example, either sign of the $\tilde{Z} \tilde{Z}$ term breaks the symmetry (either ferromagnetically or anti-ferromagnetically) in the absence of the transverse field.  However, we do not expect that this behavior is generic, for example, the $Z_N$ models we consider below do possess a gapless phase at the edge.  This is plausible given that a $U(1)$ symmetry cannot break spontaneously and that $Z_N$ mimics $U(1)$ for sufficiently large $N$.

Suppose we similarly truncate the Hamiltonian of the gauged CZX model.  Let us further suppose for simplicity that we impose the electric term as a hard constraint.  This holds even at the edge where we must modify the nature of the electric constraint term since we are now missing the vertical gauge field.  Everything still commutes and all electric strings remain tensionless. Let us first set $\tau^z = 1$ on all links.  This product state is the ground state before projection.  Clearly the boson model has exactly the same spectrum as before.  For every state of the boson sector in the gapless edge manifold we may project onto the constraint subspace to produce a physical wavefunction.  This attaches strings to the boson charges, but as we already said, these strings are tensionless.  Hence the energy is not changed upon projection and all states remain exactly degenerate.  We note in passing that the local $Z_2$ transformations look quite strange at the edge where the global symmetry is realized in a spatially extended fashion.

Now consider perturbations to the edge.  We can take symmetric operators in the boson sector that will descend to symmetric terms acting on the edge degrees of freedom e.g. of the forms we discussed above.  We then gauge these operators as before to produce gauge invariant perturbations.  Furthermore, these perturbations reduce to the usual symmetric forms discussed above when we set $\tau^z = 1$ on all links.  Such terms will not gap the edge unless they cause it to spontaneously break the gauge symmetry.  In the ungauged model we could add local terms, charged under the $Z_2$ symmetry, that would gap the edge states, but because this perturbation is charged we must attach a Wilson line to render it gauge invariant.  Since this Wilson line must end on another gauge non-invariant operator, the gauge structure forces $Z_2$ charged perturbations to come in pairs.  Each such pair will be invariant under a global $Z_2$ transformation and will be local.  If the Wilson line connects two perturbations on the boundary we obtain just another term of the type considered above.  However, the Wilson line could also stretch into the bulk.  These terms actually look dangerous, as if charge is leaking off the edge, but this is an illusion since charge cannot escape into the gapped bulk.

We have argued that in the non-fluctuating gauge field limit the gapless edge states necessarily remain.  However, there is the question of what happens when the gauge fields fluctuate.  Of course, these fluctuations are gapped so we might already guess that they are innocuous.  Generally we think of a deconfined phase as one where we can smoothly set the gauge coupling to zero.  However, what could such a fluctuating gauge field do?  It will not confine any charges by assumption.  It can certainly renormalize the edge Hamiltonian, but we have already discussed such perturbations.  Since the string tension term $h \tau^x$ flip $\tau^z$, we may analyze its effects for small $h$ by studying the boson model in a slowly varying background field.  Such a background field may not be translation invariant, but according to Ref. \cite{spt1} the edge states are robust to disorder so long as the symmetry is preserved.  Simply put, we expect that sprinkling a few $\tau^z=-1$ bonds throughout the edge will not disrupt the gapless states since the symmetry is still preserved and the edge states may tunnel through short barriers.  If we computed thermodynamic quantities, say, by summing over configurations in a path integral, then because each configuration has edge states, the gapless edge states will be visible in thermodynamics even with the gauge fluctuations.

We have so far assumed, following Ref. \cite{spt1}, that static symmetry disorder will not localize the edge states.  One piece of evidence in favor of this conclusion is the extended nature of the edge symmetry.  No matter how we group sites at the edge, we cannot reduce the symmetry to an onsite form.  Indeed, the proof of the absence of a matrix product state in Ref. \cite{spt1} does not rely on translation invariance, so in this sense the edge cannot be in a short range entangled state unless the symmetry breaks even in absence of translation symmetry.  Indeed, our argument for the lack of short ranged states compatible with the symmetry can also be generalized to translation non-invariant situations.  Nevertheless, the fate of the disordered model remains non-trivial, for example, the disordered spin-1/2 Heisenberg chain flows to an infinite randomness fixed point and while it remains highly entangled on average (hence forbidding a matrix product description), the structure of the excitation spectrum is quite unlike that of a conventional CFT or a $z\neq 1$ scale invariant theory.  Note also that for the spin-1/2 chain dimerization can immediately lead to a gapped symmetric phase, albeit one that breaks translation invariance.  Of course, it is now known that fermionic topological insulators have unusual localization properties as discussed, for example, in Ref. \cite{ti10fold}.  Clearly the effects of disorder are an interesting target for future studies.

\section{Generalizations in various dimensions}
We have given a detailed discussion of our ideas in the context of the CZX model, but they are more general.  It is possible to construct an exactly solvable boson model realizing every symmetry protected phase discussed in Ref. \cite{spt2}, and since the symmetries of these models are always defined onsite, there is no obstruction to gauging the symmetry.  Thus in general we must consider $G$ gauge theories coupled to $G$ NLSMs with topological $\theta$ terms.  We hasten to add that this may not be the most general possibility.  The models are of the Kitaev-Levin-Wen type with a non-fluctuating gauge field coupled to gapped charge matter and are exactly solvable.  Including gauge fluctuations is not expected to modify the situation given the bulk.  An interesting question arises as to what exactly these phase are?  Since they possess long range entanglement and anyonic quasiparticles, and since there are a finite number of such phases with a given number of quasiparticles (see Ref. \cite{tqc_wang,stringnet,kitaevexact}), we expect that these phases are actually of the type considered in Ref. \cite{stringnet}, at least regarding their bulk properties.  Indeed, the classification of Ref. \cite{topo_gauge_group_coho} appears to coincide with the classification in Ref. \cite{spt2}.  This expectation is confirmed in Ref. \cite{sptbraid} where it is shown that the bulk quasiparticles of the gauged CZX model correspond to those of the doubled semion model.

For the purposes of the NLSM, we have argued that the main effect of the emergent gauge structure is to prevent explicit symmetry breaking terms from appearing.  Although we discussed the case of $Z_2$, the gauging procedure and our arguments are completely general to any group $G$ with a symmetry protected short ranged phase.  The effect of gauging the $G$ symmetry is that explicit symmetry breaking operators are forbidden in the deconfined phase.  The only remaining way to avoid gapless edge states is if the edge spontaneously breaks the symmetry.  We know this behavior does not occur for continuous groups, but these tend to confine in $2+1$d without a Chern-Simons term or gapless matter.  For discrete groups, the $Z_2$ case may require fine tuning, but more generally we expect a finite parameter regime in which the edge remains gapless.  In other words, we expect the absence of symmetric relevant perturbations.  Once we gauge the discrete symmetry explicit symmetry breaking perturbations are not allowed, thus the edge will be perturbatively stable to all deformations.  We note again that the case of continuous symmetries is more complicated since these gauge theories often confine in two dimensions or are gapless in higher dimensions, furthermore, gauging a continuous symmetry on the edge can change the central charge of the edge theory potentially removing gapless degrees of freedom, where as gauging a discrete symmetry (orbifolding) does not reduce the central charge \cite{2dcft_intro}.

We can also generalize our models to three dimensions.  However, the exactly solvable models are extremely complicated, so we will not consider them explicitly.  We begin by pointing out that the group cohomology classification is different in three dimensions, for example, there is only one distinct $Z_2$ short ranged phase.  On the other hand, the groups $Z_N \times Z_N$ do have non-trivial short ranged phases.  Thus we can consider an ideal Hamiltonian with surface states protected by a $Z_N \times Z_N$ symmetry.  We comment on the surface states and their robustness below.  More generally, we can consider solvable models for any group $G$ whose group cohomology indicates non-trivial symmetry protected states in three dimensions.  We may also gauge the symmetry $G$ using an identical procedure as in two dimensions.  Given an ungauged Hamiltonian one can immediately write down a gauged Hamiltonian which is also exactly solvable.

Regarding the ungauged edge model, we still have the NLSM story involving a boundary WZW term that is supposed to render the edge gapless.  Unlike in the two dimensional case, there is no formal argument from, say, tensor network states that the surface state cannot be gapped, symmetric, and short range entangled.  The intuition from Ref. \cite{spt2} is simply that a gapped, symmetric, short range entangled phase cannot possess such a ``non-local" term in its effective action.  On a more practical level, the unusual surface symmetry appears to prevent the addition of operators that could drive the system into a ``paramagnetic" phase.  Thus we see at least three possibilities:
\begin{itemize}
\item the surface is gapped by spontaneous symmetry breaking,
\item the surface is long range entangled because it is gapless (unless we fine tune the dispersion to zero),
\item or the surface is gapped and long range entangled because it forms a topological state.
\end{itemize}
Option three was not available for a one dimensional edge. There will be a gapless phase so long as all relevant operators are forbidden by the emergent symmetry and the WZW term.  If there is no gapless phase, then again we find the intriguing possibility that the surface either breaks the symmetry or enters a topologically ordered phase.

Our direct approach to proving absence of symmetric short ranged states, as exemplified by our analysis of the CZX model above, can be generalized on a case by case basis, and we are currently working on a formal argument in three dimensions and beyond.  Although no formal proof along the lines of the matrix product state argument of Ref. \cite{spt1} has been given for higher dimensional surface states, we believe this is likely a technical challenge.  We believe the proper setting to formulate such an argument is provided by the multi-scale entanglement renormalization ansatz (MERA) (see Ref. \cite{vidalmera}).  MERA circuits of finite depth should be able to represent any gapped, symmetric, short range entangled state.  We are currently trying to better understand the action of the relevant non-onsite symmetry operators on MERA states in order to provide a proof of the conjecture in Ref. \cite{spt2}.

Let us conclude this section with a somewhat different example in three dimensions that may be divided into a ``matter" sector and a ``gauge" sector.  Suppose that the gauge sector does not experience any qualitative change in its low energy structure over the whole space while the matter sector has a domain wall separating a symmetry protected phase from a trivial phase.  We further assume that the symmetry protected phase has gapless boundary states at such a domain wall in the absence of the gauge field.  In such a situation, we generically expect that the gapless states on the domain wall will remain intact even at finite gauge coupling.

In fact, this situation is already realized experimentally, after all the physical electron is charged under the electromagnetic gauge field, and although we often treat electromagnetism as a background field, it does fluctuate in real systems. Such a realization is provided by the interface between Bi2Te3 in its topological phase and vacuum where the gauge field is the physical electromagnetic field.  Now experimentally these surface states are seen in ARPES (see Ref. \cite{arpesbi2te3}), so we must conclude that within the resolution of ARPES the gapless states remain even at finite gauge coupling.

But what is the gauge field is not free outside the material?  Returning to the case of fractionalization, a convenient example is provided by the $3+1$ dimensional topological Mott insulator proposed in Ref. \cite{topomott} and further analyzed in Ref. \cite{surf3dgauge}.  This state may be understood by decomposing the electron operator as $c_\sigma = b f_\sigma$ where $b$ is a boson that carries the electron charge and $f_\sigma$ is a fermion carrying the electron spin.  This decomposition has a $U(1)$ redundancy given by $b \rightarrow e^{i \theta} b$ and $f_\sigma \rightarrow e^{-i \theta} f_\sigma$ and hence $b$ and $f$ are coupled by a $U(1)$ gauge field.  Let us consider four distinct phases of this system based on whether the boson is in a superfluid or Mott insulating phase and whether the fermion is in a trivial or topological band insulator.  If the boson is condensed and the fermion is trivial then we obtain a trivial insulator of electrons (I), and if the boson is condensed and the fermion is non-trivial then we obtain a topological insulator for electrons (TI).  In both these phases the $U(1)$ gauge field is screened by the boson condensate and effectively disappears from the spectrum.  If the boson is in its Mott insulating phase, then the gauge field is no longer screened and we have two kinds of fractionalized phases depending on whether the $f$s form a topological (FTI) or trivial insulator (FI).

We now describe the six possible interfaces between these states.  We assume throughout that time reversal and the electromagnetic $U(1)$ are unbroken.  We know that at an I/TI interface there are gapless electronic states.  We also know, based on our discussion above, that at an FI/FTI interface there will be gapless states of the $f$ electrons.  At an I/FI interface we do not expect any surface states since the boson is simply condensing and screening the gauge field.  Similarly, at a TI/FTI interface we also do not expect to find surface states since the bandstructure of the $f$ fermion is not changing.  At a FTI/I interface we expect gapless electronic states (rather than gapless states of the $f$ electrons) since the condensed boson in the $I$ phase will, via the proximity effect, screen the $f$ fermions with bosons to produce electrons near the interface.  Finally, at an FI/TI we also expect electronic surface states.  We should point out that those interfaces where more than one thing is happening e.g. the boson is condensing and the fermion is changing its bandstructure are more sensitive to microscopic details since we can always arrange for the changes of phase to occur in microscopically different regions.  For example, the proximity effect depends exponentially on distance and hence our arguments can depend on the precise length scales of the interface.  A clean way to handle this is to approximate the interfaces in question with double interfaces where at each interface only one component (boson or fermion) changes it state e.g. I/FTI $\approx$ I/FI/FTI or I/TI/FTI.

One very important interface type we have not yet considered is the interface with vacuum.  In many respects the vacuum behaves like a trivial insulator, but there are subtle differences.  Because the vacuum contains no electron charge, the boson $b$ should certainly not be regarded as condensed in the vacuum.  On the other hand, the emergent $U(1)$ gauge structure is certainly absent in the vacuum since it emerges from the collective dynamics of electrons and there are none present.  We may refer to the vacuum as a confined phase of the gauge theory in the sense that physical electrons injected into the vacuum will not fractionalize into $b$ and $f$ particles.  So the question arises as to what surface states, if any, exist at an interface with vacuum.  We can make a few statements right away.  A vac/I interface has no protected states while a vac/TI interface does.  It is also believed that a vac/FI interface has no protected edge states, but what about a vac/FTI interface?  Suppose we model this interface as a vac/FI/FTI interface, that is, an interface where, as we emerge from the bulk, the fermions change their state before the bosons and fermions go to zero density and the gauge theory confines.  Then we will have protected surface states of the $f$ fermions coming from the FI/FTI interface; furthermore, nothing obviously singular happens as the width of the FI region shrinks to zero.  On the other hand, if we model the vac/FTI interface as a vac/TI/FTI interface then we have surface states of electrons.  In the first case the fermion first changes its bandstructure while in the second case the boson first condenses, but the second case is special because as we already noted, the boson is certainly not condensed in the vacuum.  Hence something singular is happening as the width of the TI region shrinks to zero, namely the condensate $\langle b\rangle$ which is converting the surface state of $f$ fermions into a surface state of $c$ electrons is potentially vanishing.  We can certainly have a stable surface phase where $b$ is not condensed.  If the condensate vanishes, then we might again expect to recover surface states of the $f$ fermions.  This is indeed what was found in Ref. \cite{surf3dgauge} where the boson is simply uncondensed and the gauge field is shown not to destroy the gapless surface states (although it does modify the details of the surface).

Another perspective on the question of surface states and boundary conditions is provided by the bag model of quark confinement introduced in Ref. \cite{bagmodel} (see Ref. \cite{bagreview} for an easy introduction).  This model, originally developed in the context of confinement in QCD, supposes that quarks and gluons are confined into regions called ``bags".  These bags may be roughly understood as deconfined bubbles within a larger confined vacuum.  The spectrum of quarks and gluons confined to such a bag with appropriate boundary conditions then represents an approximation to the spectrum of QCD.  Furthermore, the bag is not static but may fluctuate in spacetime leading to a variety of interesting dynamical effects.  Besides modeling hadrons, the bag model has also been applied to situations like quark matter deep inside neutron stars where a deconfined quark phase gives way to a confinded phase of hadrons as the surface of the neutron star is approached.  Below we will analyze our $U(1)$ toy model in the context of the bag model and show that within this framework there are indeed surface states at a vac/FTI interface.

Let us work in $3+1$ dimensions and consider a bag containing a Dirac fermion $\psi$ and a $U(1)$ gauge field $a_\mu$.  We will assume that the bag surface is fixed e.g. at the edge of our material and does not fluctuate.  The gauge field couples to a conserved current given by $j^\mu = \bar{\psi} \gamma^\mu \psi$.  Our conventions for the $\gamma$ matrices are
\beq
\gamma^\mu = \left[\begin{array}{cc}
               0 & \bar{\sigma}^\mu \\
               \sigma^\mu & 0
             \end{array}\right]
\eeq
with $\sigma^\mu = (1,\sigma^i)$ and $\bar{\sigma}^\mu = (1,-\sigma^i)$.  The requirement that no charge leave the bag is given by the constraint $n_\mu j^\mu = 0$ where $n$ is a unit normal to the bag boundary.  This constraint may be written explicitly as $\bar{\psi} i \gamma \cdot n \psi = 0$.  Using the fact that $(\gamma\cdot n)^+ = - \gamma \cdot n$ for spacelike $n$ and $(i \gamma \cdot n)^2 = 1$, we have the freedom to choose $i \gamma \cdot n \psi  = \pm \psi$ at the inner boundary of the bag.  This also implies that $\bar{\psi} \psi = 0$ at the bag boundary.

Now suppose that in addition to the bag boundary, the $\psi$ fermion mass also changes sign at the bag boundary.  Without the bag there would be a surface at this interface, but is this zero mode consistent with the bag model boundary conditions?  Outcome.  Let $x^3$ be the direction normal to the local bag boundary (we have in mind a very large bag).  The Dirac equation with the varying mass $m(x^3) = m \,\text{sgn}(x^3)$ is
\beq
i \gamma^\mu \partial_\mu \psi = m(x^3) \psi
\eeq
and we look for a solution with $\partial_\mu \psi = 0$ for $\mu \neq 3$.  However, we see immediately that a solution can be found if $\psi$ satisfies $i \gamma^3 \psi = \text{sgn}(x^3) \psi$ where $i \gamma^3 = i \gamma \cdot n$.  Note that if $i \gamma^3 \psi = \pm \psi$ then we have $\bar{\psi}\psi = \psi^+ \gamma^0 \psi = \bar{\psi} \pm i \gamma^3 \psi$, but this last expression satisfies
\beq
\pm \bar{\psi} i \gamma^3 \psi = \pm \psi^+ (- i \gamma^3)\gamma^0 \psi = \mp (i \gamma^3 \psi)^+ \gamma^0 \psi = - \bar{\psi} \psi
\eeq
which implies that $\bar{\psi} \psi =0$ in such a situation.  Thus the surface mode is compatible with the bag boundary conditions.

If we now include the gauge field, we must specify further boundary conditions.  The original bag model boundary conditions of Ref. \cite{bagmodel} are $n^\mu F_{\mu \nu} = 0$ which ensure that the bag is always a charge neutral object.  One heuristic argument for these boundary conditions comes from the dual superconducting model of confinement.  Just as magnetic monopoles are linearly confined in a conventional superconductor (by vortex lines), the electric charges in a dual superconductor are similarly linearly confined.  In this model the dual variables $\tilde{F}_{\mu \nu} = \frac{1}{2} \epsilon_{\mu \nu \lambda \sigma} F^{\lambda \sigma}$ satisfy conventional superconducting boundary conditions $\epsilon_{\mu \nu \lambda \sigma} n^\nu \tilde{F}^{\lambda \sigma}$ i.e. that the tangential dual electric field and normal dual magnetic field vanish at the boundary, but these conditions imply $n^\mu F_{\mu \nu}$.  In a gauge where $A^3 = 0$ we find that the condition $F_{3i} = 0$ may be simply reduced to $\partial_3 A_i = 0 $ at the boundary. Furthermore, the condition $F_{30} = 0$ reduces to $\partial_3 A^0 = 0$.  The resulting model of surface modes coupled to a $U(1)$ gauge field with these boundary conditions has been analyzed in Ref. \cite{surf3dgauge} where it was found the surface state survives (those authors choose a slightly different gauge, but we do not expect this to seriously affect the results).

We have seen that topological Mott insulators can have surface states at an interface with vacuum (provided time reversal is unbroken).  On the other hand, these states are not absolutely stable, that is they can be gapped without closing the bulk gap.  One possibility is that $b$ condenses near the surface in which case there will remain edge states but of electrons and not $f$ fermions.  Another possibility is that the $f$ fermions, which sit in a finite density metallic-like state on the surface, could spontaneously pair and condense.  In this case the surface behaves like a superconducting layer for the internal $U(1)$ gauge field.  We see that there are numerous possibilities and the complete analysis is complex.  If the electromagnetic $U(1)$ and time reversal are unbroken then we certainly have surface states of some type, but in general the surface states break up into different perturbatively stable classes.  Which surface states are realized in a given model then depends on the microscopic details of the interface.

\section{Discussion}
Here we list our main results and conclusions.
\begin{itemize}
\item{[Non-onsite symmetries]} The non-onsite nature of the symmetry prevents, in a rather direct way, the existence of short ranged symmetric ground states.
\item{[Bulk-edge correspondence]} The entanglement spectrum of a large bulk region reproduces the universal features of the physical edge spectrum e.g. it is gapless or spontaneously breaks symmetry.
\item{[Fractionalized phases]} For any short range phase in 2d or 3d protected by an onsite symmetry $G$, we can construct a stable fractionalized phase with an emergent $G$ gauge symmetry so long as the $G$ gauge theory has a deconfined phase.  We can further construct exactly solvable models for these phases.
\item{[Robust edge states]} In some cases these fractionalized phases possess perturbatively robust gapless states at a boundary between two distinct deconfined phases as well as at a vacuum boundary.
\end{itemize}

The last point deserves further comment.  We can distinguish between edge states that are totally robust, edge states that are totally robust in the presence of a symmetry, and those that are perturbatively robust.  The latter category refers to edge states that can be removed by sufficiently strong perturbations but which are described by an RG fixed point with no perturbatively relevant directions i.e. for small deformations.  The exemplar of the first category are the quantum Hall edges while the second category describe systems like topological insulators.  Our conjecture apparently can only refer to the third category since it seems that a weakly coupled Higgs-type phase at the boundary would always remove any putative boundary states.

We have we argued that by weakly gauging the symmetry in a symmetry protected phase, we can trade symmetry protection for perturbative stability.  The symmetry protected edge cannot be gapped without breaking the symmetry (protection) yet there are relevant perturbations that produce a gap e.g. those that break the symmetry.  The gauged symmetry protected edge can be gapped (no physical symmetry protects it) but it can be perturbatively stable e.g. the gauged $Z_N$ ($N>2$) case discussed above.  Since such gauged edges are not protected, the question naturally arises as to what extent they can be regarded as stand alone theories in a lower dimension without the bulk.  This questions would be interesting to study in the future, so here we only offer a few brief comments.

From the point of view discussed at the end of section 2, namely that the edge may be understood as carrying the symmetry chirally, it seems unlikely that such state could exist on its own.  For example, consider the $U(1)$ states discussed in Refs. \cite{boson_iqh1,boson_iqh2}.  These states all have an even Hall response for a background gauge field coupled to the conserved $U(1)$, but this implies that the edge current suffers an anomaly e.g. it is not conserved in general.  The explicit process which removes charge from the edge corresponds to flux insertion in the bulk.   The existence of the Hall response for the background $U(1)$ gauge field implies that letting the $U(1)$ gauge field fluctuate leads to a gapped topological state in the bulk (confinement of pure $U(1)$ gauge theory is avoided).  However, certainly the gauged chiral edge theory is inconsistent without the bulk.  Note that the bulk theory is somewhat similar to the chiral spin state discussed in Refs. \cite{chiral_spin,edge_chiral_spin}.  In particular, in Ref. \cite{edge_chiral_spin} it is argued that the degrees of freedom associated with the $U(1)$ edge modes are removed by the gauging process, and since we have no additional $SU(2)$ symmetry it appears that the chiral edge states originally carrying the $U(1)$ charge are completely removed from the spectrum.  This leaves the intriguing possibility that the remaining non-charged modes with the opposite chirality could survive on the edge.  This possibility deserves further study.

The phases we have constructed are, at least as far as the bulk is concerned, identical to previously known long range entangled phases as argued in Ref. \cite{sptbraid}.  This connection is important because the models constructed in Ref. \cite{stringnet} are not generally expected to have protected edge states, so this correspondence agrees with our expectation that the gauged theories we consider do not have protected boundary states. Nevertheless, perturbatively stable gapless boundary modes are quite interesting in their own right.  We also reiterate the point that while the bulk may be identical, there appears to be subtle questions related to boundary conditions especially in the duality arguments.  It is possible that while there is only one bulk doubled semion phase, there are multiple renormalization group stable boundary conditions, some of which support gapless edge states.  General considerations along these lines, although for gapped boundaries, have recently been given in Ref. \cite{kkdomwalls}.  We should also point out that some of the models we have constructed are amenable to study via quantum Monte Carlo and hence the question of edge states should be answerable numerically for those models.

\textit{Acknowledgments}
We thank Xie Chen and Jay Sau for many valuable conversations.  We also thank John McGreevy, Xiao-Gang Wen, Mike Mulligan, and Zheng-Cheng Gu for helpful conversations and Perimeter Institute for hospitality in the early stages of this work.  BGS is supported by a Simons Fellowship through Harvard University.

\bibliography{gczx_paper}

\end{document}